\documentclass[pra, amsmath,amssymb,preprint]{revtex4-2}

\usepackage{graphicx}
\usepackage{dcolumn} 
\usepackage{bm}
\usepackage[utf8]{inputenc}
\usepackage[T1]{fontenc}
\usepackage{mathptmx}

\usepackage{color} 
\usepackage{xcolor}
\usepackage[caption=false]{subfig} 

\definecolor{myblue}{gray}{0.5}

\long\def\comment#1{}

 \newcommand{\rhob}{\rho_{i\sigma}}

\begin{document}

\title{Fermi-L\"owdin orbital self-interaction
correction using the optimized effective potential method  within the Krieger-Li-Iafrate approximation}

\author{Carlos M. Diaz}
\email{cmdiaz6@miners.utep.edu}
\affiliation{Department of Physics, University of Texas at El Paso, El Paso, Texas 79968, USA}

\author{Tunna Baruah}
\email{tbaruah@utep.edu}
 \affiliation{Department of Physics, University of Texas at El Paso, El Paso, Texas 79968, USA}

\author{Rajendra R. Zope}
\email{rzope@utep.edu}
 \affiliation{Department of Physics, University of Texas at El Paso, El Paso, Texas 79968, USA}%

\date{\today}

\begin{abstract}
 Perdew-Zunger self-interaction correction (PZ-SIC) offers a route to remove 
 self-interaction errors on an orbital-by-orbital basis.
 A recent formulation  of PZ-SIC by Pederson, Ruzsinszky and Perdew proposes restricting 
 the unitary transformation to localized orbitals called Fermi-L\"owdin orbitals.
 This formulation, called the FLOSIC method, simplifies 
 PZ-SIC calculations and was implemented self-consistently using a Jacobi-like
 (FLOSIC-Jacobi)
 iteration scheme. In this work we implement the FLOSIC approach  using the 
 Krieger-Li-Iafrate (KLI) approximation to the optimized effective potential (OEP).
 We compare the results of present FLOSIC-KLI approach with FLOSIC-Jacobi 
 scheme for atomic energies, atomization energies, ionization energies, 
 barrier heights,  polarizability of chains of hydrogen molecules etc.
 to validate the FLOSIC-KLI approach. The FLOSIC-KLI approach, 
 which is within the realm of Kohn-Sham theory, predicts
 smaller energy gaps between frontier orbitals due to the lowering 
 of eigenvalues of the lowest unoccupied orbitals. Results show that 
 atomic energies, atomization energies, ionization energy as an 
 absolute of highest occupied orbital eigenvalue, and polarizability of 
 chains of hydrogen molecules between the two methods agree within 2\%.
 Finally the FLOSIC-KLI approach is used to determine the vertical 
 ionization energies of water clusters.

\end{abstract}

\maketitle

\section{Introduction}
The Kohn-Sham (KS) formulation of the density functional theory (DFT) is
an exact theory widely used 
in chemical physics, materials science and condensed matter physics\cite{jones2015density}. 
Its practical usage requires  approximations to  the exchange-correlation functional
whose accuracy and complexity determines the accuracy and efficiency of the study.
As there is no systematic way to improve upon the accuracy of exchange-correlation 
approximations, a large number of density functional approximations (DFAs) have been
been proposed\cite{perdew2001jacob,MARQUES20122272}.  Practically, all these 
functionals suffer from self-interaction-error (SIE) which has restricted the 
universal application of DFT. The SIE has been 
attributed to the problem of excessive delocalization of electrons, low 
reaction barrier heights, overestimation of eigenvalues of occupied orbitals, 
overestimation of polarizabilities of 
molecular chains, underestimation of band gaps,  etc.
In KS-DFT, when the exchange-correlation functional is approximated, the 
self-Coulomb energy included in the expression of Coulomb energy does not 
get fully cancelled by the self-exchange in the approximate exchange-correlation 
functional. The residual left is the self-interaction energy. For example, 
\textcolor{black}{
for the hydrogen atom or one electron densities $\rhob$ of spin $\sigma$  the sum of Coulomb energy $E_H$
and exchange-correlation $E_{xc}$ is 
\begin{equation}
   E_H  + E_{xc} = \frac{1}{2}\iint d^3r \, d^3r' \frac{\rhob(\vec{r})\rhob(\vec{r}')}{|\vec{r}-\vec{r}'|}
   +\, E_{xc} [\rhob] = \delta. 
\end{equation}
}
For the exact functional $\delta=0$. For 
approximate functionals,  $\delta$ is non-zero and represents 
 the self-interaction error for that functional for the one-electron density.

Several approaches have been proposed to remove the SIE explicitly 
\cite{lindgren1971statistical, PhysRevA.15.2135, perdew1982density, lundin2001novel, doi:10.1063/1.2403848,
gidopoulos2012constraining,PhysRevB.76.033102, doi:10.1002/jcc.10279, borghi2014koopmans, 
doi:10.1063/1.2204599, doi:10.1063/1.5129533, 
doi:10.1063/1.4866996}. 
Early approaches\cite{lindgren1971statistical,PhysRevA.15.2135} 
 used  orbital-wise schemes to eliminate the SIE but used functionals related to Slater's X$\alpha$ method \cite{Slater1951}.
More common approaches that mitigate SIE include
hybrid functionals, which mix Hartree-Fock exchange using various 
criteria\cite{doi:10.1063/1.464304,iikura2001long,jaramillo2003local,baer2010tuned}.
A large 
literature on the hybrid functionals that were introduced by Becke\cite{doi:10.1063/1.464304} exist,
but these approaches are not entirely self-interaction free and are challenging 
for extended systems.
\subsection{ Perdew-Zunger SIC}
 In 1981, Perdew and Zunger (PZ)\cite{PhysRevB.23.5048} proposed a method to remove the 
one-electron SIE in an orbital-wise fashion. This method is the most common approach to 
explicitly remove the SIE. 
PZ-SIC provides the exact cancellation for 
one-electron self-interaction (SI), but not necessarily for many-electron SI\cite{doi:10.1063/1.2566637}.
 In the PZ-SIC method, \cite{PhysRevB.23.5048}  
 the orbital-wise SIC to the total energy is
\begin{equation} \label{eq:SICA}
 E^{SIC}=-\sum_{i\sigma}^{N_{occ}} \left ( U[\rho_{i\sigma} ]
 + E_{xc}^{DFA} [\rho_{i\sigma},0] \right ).
\end{equation}
\textcolor{black}{
Here, $U[\rho_{i\sigma}]$ and $E_{xc}^{DFA} [\rho_{i\sigma},0 ]$ 
are the Coulomb and exchange-correlation energy of the $i^{th}$ occupied orbital,
$\sigma$ is the spin index, $N_{occ}$ is the number of occupied orbitals, and 
$\rho_{i\sigma}$ is the orbital electron density.
}
It is obvious from 
Eq. (\ref{eq:SICA}) that the PZ-SIC corrections make the DFA exact for any one-electron density. 
\textcolor{black}{
The SIC should vanish for the exact functional. It is unclear if PZ-SIC satisfies 
this condition.
The exact functional is valid 
only for ground state densities 
while the SIC using the PZ-SIC method is obtained on an orbital-by-orbital basis, that is, using  orbital densities which are noded\cite{hofmann2012self}.
}
\textcolor{black}{
The total energy with the PZ-SIC method is given by $E =E^{KS}+E^{SIC}$.
In atomic units, $E^{KS}$ is given by
}
\begin{multline} \label{eq:Eks}
    E^{KS}=\sum_{i\sigma}  \langle \psi_{i\sigma}|-\frac{\nabla^2}{2}|\psi_{i\sigma} \rangle 
    + \int d^3r \, \rho(\vec{r})v_{ext}(\vec{r})  \\
    +\frac{1}{2}\iint d^3r \, d^3r' \frac{\rho(\vec{r})\rho(\vec{r}')}{|\vec{r}-\vec{r}'|}
    +E_{xc}[\rho_{\uparrow},\rho_{\downarrow}].
\end{multline} 
Here, $v_{ext}$ is the external potential and $\rho = \rho_{\uparrow} + \rho_{\downarrow}
=  \sum_{\sigma} \rho_{\sigma} = \sum_{i,\sigma} f_{i\sigma} \vert \psi_{i\sigma}\vert^2 $ is the electron density, where
$f_{i\sigma}$ is the occupation of the $\psi_{i\sigma}$ orbital.
\textcolor{black}{Atomic units are used throughout this article unless specified explicitly.}

The SI corrected potential seen by an electron in the $i^{th}$
orbital in the PZ-SIC method is 
\begin{eqnarray}
\label{eq:Veff}
v_{eff}^{i\sigma}(\vec{r}) & =  & v_{ext}  (\vec{r})
+ 
\int d^3r'\, \frac{\rho(\vec{r}')} {\vert \vec{r}-\vec{r}'\vert}  +  
v_{xc}^{\sigma} (\vec{r}) \nonumber \\
& - &
 \left \{ \int d^3r' \,
 \frac{\rho_{i\sigma} (\vec{r}')} {\vert \vec{r}-\vec{r}'\vert}  +  
v_{xc}^{i\sigma} (\vec{r})
 \right \}.
\end{eqnarray} 
the second term is the Coulomb 
potential due to the electrons and 
$v_{xc}$ is the exchange-correlation potential (of DFA). The last two terms in the curly bracket 
constitute the SIC potential for the $i^{th}$ orbital $v_{i\sigma}^{SIC} = -\{v_C^{i\sigma}+v_{xc}^{i\sigma}\}$, 
composed of the self-Coulomb and self-exchange-correlation potentials. 
Unlike in the standard KS equations, the potential in  Eq.~(\ref{eq:Veff})  is orbital dependent. 
This orbital dependence complicates the solution of one-electron equations. 
For atoms where the KS orbitals are localized, PZ-SIC provides finite 
SIC. However, the method is not size extensive if the KS orbitals are used. The 
Kohn-Sham orbitals are delocalized for a system made up of a
 collection of atoms with large separation between them.
These delocalized KS orbitals  give vanishing SIC 
 correction if used in the PZ-SIC method. For extended systems 
the delocalized KS orbitals are  
 normalized over the entire volume of the solid and hence 
 orbital-dependent quantities in Eq. (\ref{eq:SICA}) approaches zero  for  such systems.
 The SIC can be made 
 size extensive by using localized orbitals, which can be obtained
 from KS orbitals by unitary transformation.
 Pederson, Heaton, and Lin implemented such a SIC scheme 
 and demonstrated the first PZ-SIC calculation for molecules\cite{doi:10.1063/1.448266}. 
 In the 1980s,  Lin's group at Wisconsin used a localization approach to
 implement the PZ-SIC method \cite{PhysRevB.28.5992,doi:10.1063/1.446959,doi:10.1063/1.448266,Pederson1988}.
 The orbital-dependent Coulomb and exchange-correlation energies 
 and potentials in Eq. (\ref{eq:Veff}) are computed using local orbitals.
 The localization approach by Pederson and coworkers 
requires that the local orbitals that minimize total energy 
must satisfy  Pederson's localization equations
given below. 
\begin{equation}
\langle \phi_i \vert  H_i -  H_j \vert  \phi_j \rangle = 
\lambda_{ji}^i - \lambda_{ij}^j = 0.
\end{equation}
Here $H_i$ is the orbital dependent Hamiltonian,
$\phi$ are the localized orbitals obtained by unitary 
transformation of the KS orbitals $\psi$, and $\lambda$ are the 
Lagrangian multiplier introduced to maintain the orthogonality
constraint. When the total energy is at variational
minimum the Lagrangian multiplier matrix is symmetric.

The variational minimization of PZ-SIC energy requires satisfying 
$N(N-1)/2$ localization equations where $N$ is the number of 
occupied orbitals.
In 2014, Pederson and coworkers used L\"owdin orthogonalized Fermi-orbitals (FLOs)
in the PZ-SIC method.
The PZ-SIC using FLOs reduces the number of unknown
parameters needed to describe the unitary transformation
and reduce the number of constraints  from $~N^2$ to $3N$.
\textcolor{black}{Before closing this section we note that a localizing transformation can also 
be incorporated in the Kohn-Sham formalism using the OEP method 
as shown by K\"orzd\"orfer and coworkers\cite{Korzdorfer2008}. This 
generalized OEP method is also 
invariant under unitary transformation of the orbitals.
}
 Below we briefly describe the details of 
the PZ-SIC using FLOs.

\comment{
The problem can nevertheless be solved
within the framework of the optimized effective
potential (OEP) method [19,20]. The OEP method
yields an exchange-correlation potential in the
form of a local and multiplicative operator, which
is the same for all the orbitals
}

\subsection{Fermi-L\"owdin orbital SIC (FLO-SIC)} \label{sec:FLOSIC}
 Recently, Pederson, Ruzsinszky, and Perdew \cite{doi:10.1063/1.4869581} introduced a unitary invariant 
     implementation of PZ-SIC using Fermi-L\"owdin orbitals \cite{Luken1982,Luken1984} 
     called the FLO-SIC method. FLO-SIC has been used interchangeably with PZ-SIC earlier, but 
     FLOs can also be used in other variants of SIC including OSIC\cite{doi:10.1063/1.2176608}, 
     SOSIC\cite{doi:10.1063/5.0004738}, and recently introduced
     local scaling SIC\cite{doi:10.1063/1.5129533} methods.
FLO-SIC makes use of localized Fermi orbitals (FOs) $F_{i\sigma}$ which are defined by the transformation of KS orbitals as
\begin{equation}
    F_{i\sigma}(\vec{r})=\frac{\sum_\alpha \psi_{\alpha\sigma}^{*}(\vec{a}_{i\sigma})\psi_{\alpha\sigma}(\vec{r})}{ \sqrt{\sum_\alpha |\psi_{\alpha\sigma}(\vec{a}_{i\sigma})|^2}}.
\end{equation}
Here, $\vec{a}_{i\sigma}$ are points in space called Fermi-orbital descriptors (FODs). Neglecting the spin index,
the above equation can be rewritten as
\begin{equation}
    F_i(\vec{r})=\sum_{\alpha}^{N_{occ}}F_{i\alpha}\psi_{\alpha}=\frac{\rho(\vec{a}_i,\vec{r})}{\sqrt{\rho(\vec{a}_i)}},
\end{equation}
where the transformation matrix $F_{i\alpha}$ is defined as
\begin{equation} \label{eq:Tij}
    F_{i\alpha}=\frac{\psi_{\alpha}^*(\vec{a}_i )}{\sqrt{\rho(\vec{a}_i)}}.
\end{equation}
The FOs are normalized but are not orthogonal. They are orthogonalized using the 
L\"owdin orthogonalization method 
to generate the Fermi-L\"owdin orbitals (FLOs) $\phi_{i\sigma}$. 
Optimal FOD positions are found using gradients of the energy with respect to FOD positions using 
minimization procedures analogous to geometry optimizations \cite{doi:10.1063/1.4907592, PEDERSON2015153}. A number of studies 
have been conducted using the FLOSIC method \cite{doi:10.1080/00268976.2016.1225992,PhysRevA.95.052505,  doi:10.1063/1.4996498, kao2017role, doi:10.1063/1.4947042, doi:10.1021/acs.jpca.8b09940, doi:10.1063/1.5050809, doi:10.1063/1.5125205, doi:10.1002/jcc.26008, C9CP06106A,doi:10.1002/jcc.25767,doi:10.1063/1.5050809,doi:10.1063/1.4996498,Jackson_2019,PhysRevA.100.012505, doi:10.1021/acs.jpca.8b09940, doi:10.1063/1.5125205,doi:10.1063/1.5125205, doi:10.1063/1.5087065, doi:10.1063/1.5129533, doi:10.1021/acs.jctc.8b00344, Sharkas11283, Jackson_2019, FLOSICSCANpaper, doi:10.1063/5.0004738, doi:10.1002/jcc.25586, doi:10.1002/jcc.25586, schwalbe2019pyflosic,romero2020local,diaz2021implementation,kshir2021}.

\subsection{Self-consistency in FLO-SIC}
Two routes have been used to implement orbital dependent functionals. The first one 
is the generalized Kohn-Sham scheme\cite{seidl1996generalized} that is widely used to 
implement hybrid functionals which contain orbital dependent Hartree-Fock exchange. 
This approach lies outside of the traditional Kohn-Sham scheme with multiplicative effective 
potentials. Within the Kohn-Sham scheme,  orbital-dependent functionals are implemented
using  the optimized effective potential (OEP) method \cite{Sharp1953,Talman1976}.

The PZ-SIC method can also been implemented using the OEP method.
In the OEP method total energy is minimized 
with respect to a local-multiplicative potential\cite{Sharp1953,Talman1976}.
This results in integral equations that are very complex 
and computationally demanding to solve. Typically 
the OEP solution is obtained using simplifications 
proposed by the  Krieger, Li, and Iafrate  (KLI)\cite{PhysRevA.46.5453}.
A few implementations of the PZ-SIC method using the KLI-OEP
have been reported\cite{doi:10.1063/1.481421,doi:10.1063/1.1370527,doi:10.1021/jp014184v,Chu2001,PhysRevA.46.5453,legrand2002comparison,Korzdorfer2008}.
For more details about the OEP-PZ-SIC method and its comparison to non-OEP approach
we refer an interested reader to  Ref. \onlinecite{Korzdorfer2008}.

Previous implementations of self-consistent FLOSIC  used an approach related to 
Jacobi rotations\cite{PhysRevA.95.052505}.  
In this approach, an approximate Hamiltonian is first constructed as
\begin{equation}
    \tilde{H}_{mn\sigma}=\langle \phi_{m\sigma} | H_\sigma^{KS} + v_{i\sigma}^{SIC} | \phi_{n\sigma} \rangle
\end{equation}
where $H_\sigma^{KS}$ is the traditional KS Hamiltonian.
(See Ref. \onlinecite{PhysRevA.95.052505}  for more details.)
The FLOs and the unoccupied virtual orbitals are made orthogonal through pairwise Jacobi rotations
which are carried out iteratively until the matrix elements for the $i^{th}$ orbital Hamiltonian between $\phi_i$ and a virtual orbital vanishes.
Alternative schemes such as a unified Hamiltonian \cite{PhysRevB.28.5992,PhysRevB.28.5992,doi:10.1021/ct500637x} and a generalized-Slater scheme in real space \cite{diaz2021implementation}
have also been used.

The purpose of this work is to introduce self-consistency in the FLO-SIC method 
using the OEP-KLI approximation. We  refer to this implementation as FLOSIC-KLI. We compare the
results obtained using FLOSIC-KLI for large number of properties 
against the 
Jacobi-rotation approach to self-consistency (FLOSIC-Jacobi) as well as 
to the experimental
values. We also use the present implementation to study the vertical ionization 
energies of water clusters containing 20 to 30 water molecules.
In Section \ref{sec:flosickliequations} we describe the FLOSIC-KLI equations.
In Section \ref{sec:results} we present results for atomic energies and highest occupied 
orbital (HOO) eigenvalues as well as total energies and atomization energies of molecules 
and compare against the  self-consistent FLOSIC-Jacobi approach as implemented in the FLOSIC code.

\section{Theory and computational details}
\subsection{FLOSIC-KLI equations}\label{sec:flosickliequations}
The present implementation of PZ-SIC using FLOSIC-KLI closely follows that of Patchkovskii, Autschbach, 
and Ziegler\cite{doi:10.1063/1.1370527}.
In the KLI approximation, the orbital dependent potential of the PZ-SIC Equation (Eq. [\ref{eq:Veff}]) is replaced by
\begin{equation}
    \label{eq:kli-veff}
    v_{eff}^\sigma(\vec{r})=v_{ext}(\vec{r})+ \int d^3 r' \frac{\rho(\vec{r'})}{|\vec{r}-\vec{r'}|} +v_{xc}^\sigma(\vec{r}) +v_{xc,\sigma}^{KLI}(\vec{r})
\end{equation}
The KLI contribution to the potential is given by the equations
\begin{equation}
    \label{eq:vkli}
    v_{xc,\sigma}^{KLI}(\vec{r})=v_{xc,\sigma}^S(\vec{r}) + \sum_{i=1}^{N_\sigma} \frac{\tilde{\rho}_{i\sigma}(\vec{r})}{\rho_\sigma(\vec{r})} (x_{i\sigma}-C_\sigma)
\end{equation}

\begin{equation}
    \tilde{\rho}_{i\sigma}(\vec{r}) = f_{i\sigma}|\phi_{i\sigma}(\vec{r})|^2.
\end{equation}
In present formulation, $\phi_{i\sigma}$ are the FLOs (localized orbitals) described in section \ref{sec:FLOSIC}.
It has been found that using $\phi_{i\sigma}$ as Kohn-Sham orbitals leads to poor results\cite{doi:10.1021/jp014184v,doi:10.1063/1.481421}. 
The leading contribution to the KLI potential is the density-weighted average of the orbital 
SIC potentials, $v_{xc,\sigma}^S$. This term is similar to the Slater approximation\cite{Slater1951}
to the average exchange potential and is given as
\begin{equation}
    \label{eq:slater}
    v_{xc,\sigma}^S(\vec{r})=\sum_{i=1}^{N_\sigma} \frac{\tilde{\rho}_{i\sigma}(\vec{r})}{\rho_\sigma(\vec{r})} v_{i\sigma}^{SIC}(\vec{r}).
\end{equation}

The second term in Eq. (\ref{eq:vkli}) allows a per-orbital shift in potentials due to the $x_{i\sigma}-C_\sigma$ factor.
The magnitudes of the shifts can be determined by enforcing a constraint that the interaction energy between 
a given localized electron and the shifted SIC potential, $v_{i\sigma}^{SIC}(\vec{r}) + x_{i\sigma} - C_\sigma$,
equals the energy of the electron in the average potential:

\begin{equation}
    \int (v_{i\sigma}^{SIC}(\vec{r}) +x_{i\sigma} -C_\sigma) \rho_{i\sigma}(\vec{r}) d\vec{r} = 
    \int v_{xc,\sigma}^{KLI}(\vec{r})\rho_{i\sigma}(\vec{r})d\vec{r}
\end{equation}
Substituting $V_{xc,\sigma}^{KLI}$ from Eq. (\ref{eq:vkli}) results in a system of linear equations for $x_{i\sigma}$: 
\begin{equation}
    \sum_{j=1}^{N_\sigma}(\delta_{ij}f_{i\sigma}-M_{ij}^\sigma)x_{j\sigma} = \overline{v}_{i\sigma}^{S}-\overline{v}_{i\sigma}^{SIC}, i=1,...,N_\sigma 
\end{equation}
where
\begin{equation}
    \label{eq:kli-M}
    M_{ij}^\sigma = \int \frac{ 
    \rho_{i\sigma}(\vec{r}) \rho_{j\sigma}(\vec{r})}
    {\rho_\sigma (\vec{r})} d\vec{r},
\end{equation}
\begin{equation}
    \label{eq:kli-vbarS}
    \overline{v}_{i\sigma}^{S}= \int
    \rho_{i\sigma}(\vec{r}) v_{xc,\sigma}^{S}(\vec{r}) d\vec{r},
\end{equation}
\begin{equation}
    \label{eq:kli-vbarSIC}
    \overline{v}_{i\sigma}^{SIC}= \int
    \rho_{i\sigma}(\vec{r}) v_{i\sigma}^{SIC}(\vec{r}) d\vec{r}.
\end{equation}
From Eqs. (\ref{eq:slater}) and (\ref{eq:kli-M}-\ref{eq:kli-vbarSIC}) it follows

\begin{equation}
    \sum_{i=1}^{N_\sigma}M_{ij}^\sigma = 1,
\end{equation}
\begin{equation}
    \sum_{i=1}^{N_\sigma} 
    (\overline{v}_{i\sigma}^S - \overline{v}_{i\sigma}^{SIC}) = 0
\end{equation}

\begin{figure*}
        \subfloat[FLOSIC-Jacobi]{ \label{subfig:scf-jacobi}
            \includegraphics[height=0.4\paperheight]{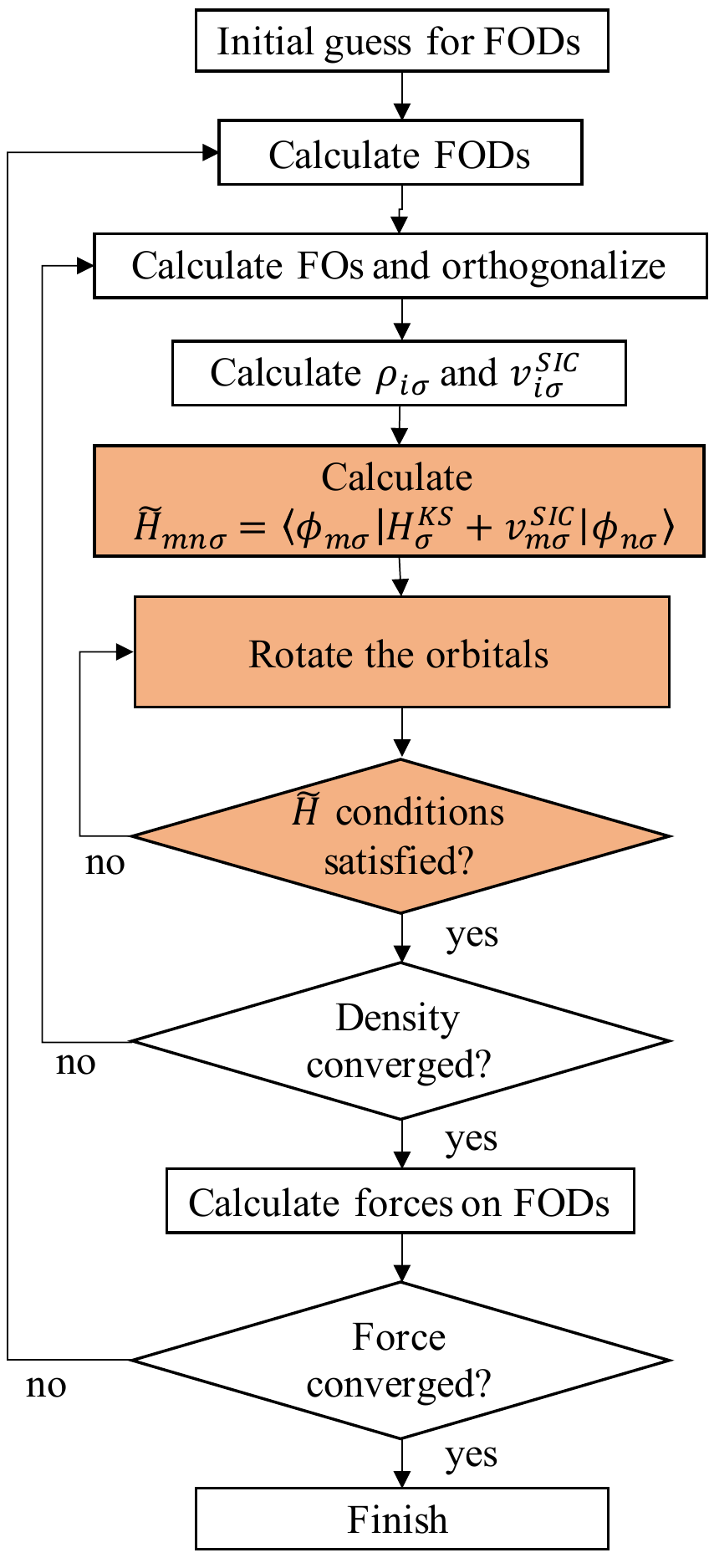}
        }
        \hspace{30 mm}
        \subfloat[FLOSIC-KLI]{ \label{subfig:scf-kli}
            \includegraphics[height=0.4\paperheight]{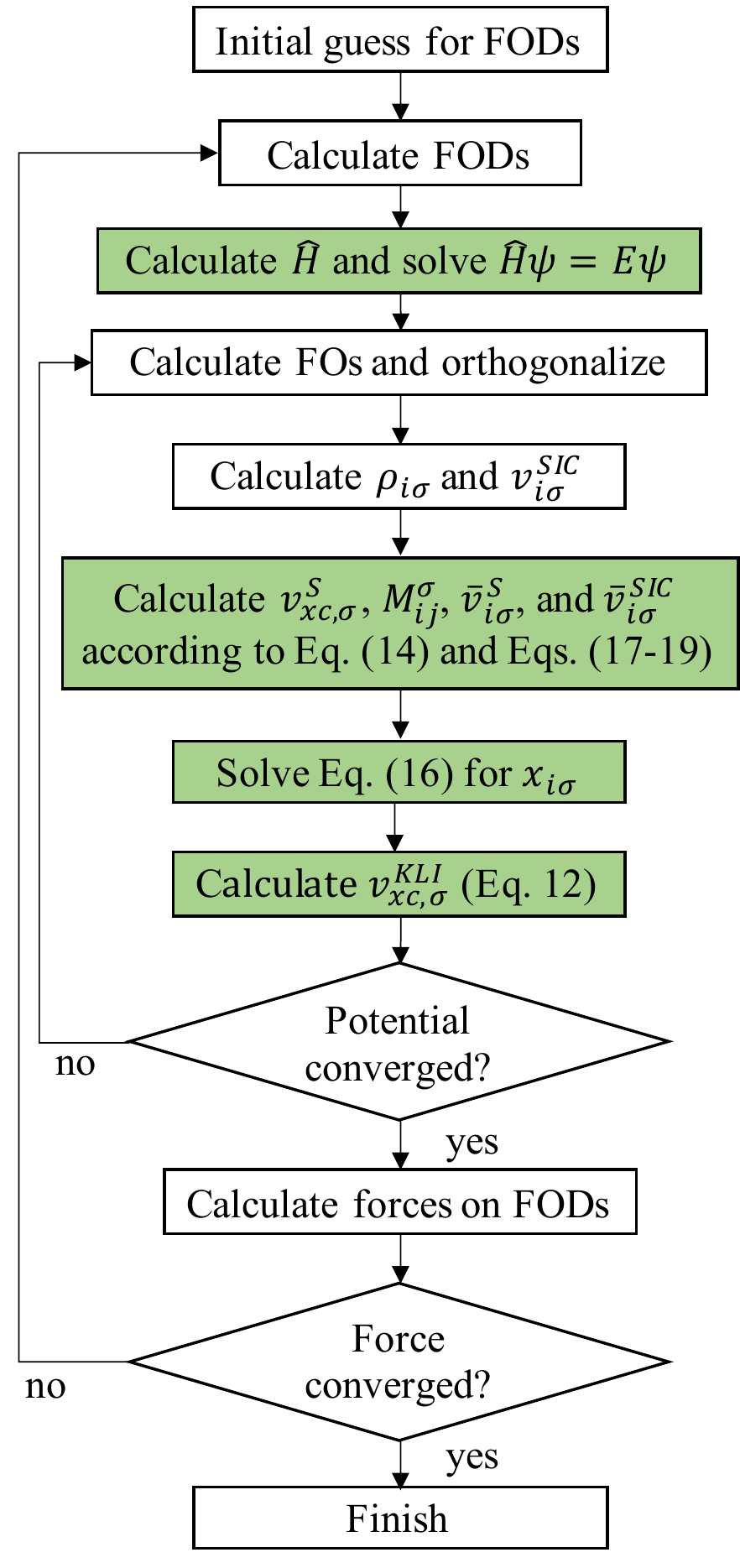}
        }
    \caption{SCF diagrams of FLOSIC-Jacobi and FLOSIC-KLI schemes. Differences highlighted in red for FLOSIC-Jacobi and green for FLOSIC-KLI.}
    \label{fig:scf}
\end{figure*}

In the original KLI approach, the values of the coefficients $x_{i\sigma}$ are chosen to satisfy

\begin{equation}
    v_{xc}^{KLI}(\vec{r}) = v_{xc,\sigma}^S(\vec{r}) + 
    \sum_{i=1}^{N_\sigma} \frac{\rho_{i\sigma}(\vec{r})}{\rho_\sigma(\vec{r})}
    (\overline{v}_{xc,i\sigma}^{KLI}-\overline{v}_{i\sigma}^{SIC})
\end{equation}
where
\begin{equation}
    \overline{v}_{xc,i\sigma}^{KLI}(\vec{r}) = 
    \int \rho_{i\sigma}(\vec{r}) v_{xc,\sigma}^{KLI}(\vec{r}) d\vec{r}.
\end{equation}

In the limit as $r\rightarrow\infty$, $\rho_\sigma$ can be expected to be dominated 
by the highest occupied molecular orbital (HOMO), $\rho_{\sigma}^{HOMO}$.
In this limit, it follows that
\begin{equation}
    v_{xc,\sigma}^{DFA}(\vec{r}) +v_{xc}^{KLI} \rightarrow
    -\frac{1}{r} +x_{\sigma}^{ HOMO}-x_\sigma.
\end{equation}

Eq. (\ref{eq:vkli}) is identical to the KLI-OEP expression if $C_\sigma$ is chosen as $C_{\sigma}=x_{\sigma}^{HOMO}$. 
For other choices of $C_\sigma$, the potentials differ by a constant. Patchkovskii et al. \cite{doi:10.1063/1.1370527} note difficulties in defining the HOMO in molecular calculations and find a choice of $C_\sigma=min(x_{i\sigma})$ to give favorable convergence properties. 
In our calculations, we find using $C_\sigma=max(x_{i\sigma})$ to give orbital energies comparable to original FLOSIC-Jacobi calculations and favorable convergence for most systems tested. 
Two exceptions were the atomic cases of lithium and sodium, where calculations 
failed to converge. In these cases, total energies 
were calculated using $C_\sigma=min(x_{i\sigma})$.
For the two problematic cases of lithium and sodium, calculations can be converged
by fixing the orbital occupation. 
This  gives the same total energies as by choosing $C_\sigma=min(x_{i\sigma})$,
but in these cases the lowest unoccupied molecular orbital (LUMO) energy is brought lower than the HOMO, which is of opposite spin.
Since orbital eigenvalues are affected by the choice of $C_\sigma$, the HOMO energies for lithium and sodium are not included in errors reported in section \ref{sec:results}. 
The steps to solve FLOSIC-KLI equations self-consistently and the difference 
of the FLOSIC-KLI implementation with FLOSIC-Jacobi scheme are illustrated in Fig. \ref{fig:scf}.

\subsection{Computational details}\label{sec:computational_details}
All of the results presented in this manuscript are calculated with the FLOSIC code\cite{FLOSICcode,FLOSICcodep},
which is based on the UTEP version of the  NRLMOL electronic structure code\cite{PhysRevB.41.7453,PhysRevB.42.3276}.
The FLOSIC code inherits the optimized Gaussian basis sets of NRLMOL\cite{PhysRevA.60.2840} 
and an accurate numerical integration grid scheme \cite{PhysRevB.41.7453}. 
The SIC calculations require a finer mesh as orbital densities are involved in calculation of orbital-dependent potentials. 
A default NRLMOL mesh for FLOSIC calculation, on average, has 25,000 grid points per atom. 
This results in integration of charge density that is accurate to the order of $10^{-8}e$.
All calculations use the Perdew, Burke \& Ernzerhof (PBE) exchange-correlation functional\cite{PhysRevLett.77.3865} 
except for the water clusters.  Water clusters calculations were performed using PBE as well 
as the local spin density approximation (LSDA).
For the LSDA \textcolor{black}{correlation, the} Perdew-Wang parameterization\cite{PhysRevB.45.13244} was used.
A self-consistency convergence tolerance of $10^{-6}$ Ha in the total energy was used for all calculations.

FLOSIC calculations require an initial set of trial FOD positions  We use previously reported PBE-optimized FOD positions. 
These FOD positions were optimized by minimizing the FOD forces\cite{doi:10.1063/1.4907592} until the convergence criteria of $10^{-6}$ Ha 
on the FLOSIC total energy was met. FOD positions were not re-optimized for KLI calculations,
\textcolor{black}{ except for the calculations on hydrogen chains in section \ref{sec:h2-chains}. 
  We note that this is an additional approximation. The FOD positions depend on the choice 
  of the Hamiltonian and the exchange-correlation approximation. We have examined the effect
  of this approximation by re-optimizing the FODs for atomic systems within the FLOSIC-KLI scheme.
  We find that the optimization results in 0.36\% change (0.58 milli-Hartree) in the mean absolute error (MAE) compared to experiment, in each case bringing the results to better agreement with the FLOSIC-Jacobi results.
  The largest observed change was a 3 milli-Hartree lowering in the case of the fluorine atom, bringing it within 3 milli-Hartree of the optimized FLOSIC-Jacobi result.
}
We refer to calculations using the Jacobi-rotation approach to self-consistency as FLOSIC-Jacobi 
and calculations using the KLI approximation as FLOSIC-KLI.   A subset of calculations 
were obtained using only a
leading term of the KLI approximation (Eq. [\ref{eq:slater}]) which we refer to as FLOSIC-Slater.
The FOD positions for the water clusters were obtained using the fodMC code\cite{Schwalbe2019}.

\subsection{ADSIC guess}
  The iterative solution of KS or PZ-SIC equations requires an initial guess to start the SCF 
  cycle. Several choices of initial guess exist. Since its inception in late 80s,
  the NRLMOL code  (on which the FLOSIC code is based) uses a linear superposition of 
  atomic potentials  (SAP) as 
  an initial guess. The atomic potentials are generated on the fly and a least square fit 
  is used to construct initial potentials for molecular systems. Our experience
  is that this choice has worked well for wide variety of systems.
  Recently, Lehtola\cite{lehtola2019assessment} has reviewed the performance of 
  various choices for initial guess to 
  initialize  the SCF cycle and has concluded that SAP on average performs better
  than other choices. Typically in FLOSIC calculations we either start from SAP 
  or from the converged DFA (SIC-uncorrected)  KS orbitals. This has worked well but 
  there are cases where starting DFA KS density can have incorrect character, for 
  example when molecules are in dissociation limits. In such case self-consistent
  FLOSIC calculations can take longer to converge. 
  An alternative if not better initial SAP  for SIC calculations 
  can be generated from the self-interaction 
  corrected atomic potentials using a suitable SIC method. We construct 
  the SAP using a simple average density SIC (ADSIC) scheme\cite{legrand2002comparison,ciofini2003mean},
  which is a generalization of the
  Fermi and Amaldi\cite{Fermi1934} method. OEP-KLI-SIC can also be used but we have 
  chosen ADSIC due to its simplicity. 
  The KS effective potential in ADSIC exhibits the correct $-1/r$ asymptotic.
In ADSIC, the self-interaction corrections to the Coulomb and exchange-correlation potential
are given by 
\begin{equation}
    V_C^{ADSIC}=V_C[\rho]-V_C[\frac{\rho}{N_e}] = V_C \frac{N_e-1}{N_e},
\end{equation}
and 
\begin{equation}
    V_{xc}^{ADSIC}=V_{xc}[\rho]-V_{xc}[\frac{\rho}{N_e}].
\end{equation}
Here, $N_e$ is the number of electrons. This correction can become very small for systems with a large 
number of electrons, but here we are using it only to generate atomic potentials.
In general, we have found that using superposition of ADSIC atomic potentials as an initial guess
in the self-consistent FLOSIC calculations usually, but not always, 
requires fewer iterations to converge than starting from SAP from DFAs or starting from 
the converged DFA orbitals. 

\subsection{KLI implementation/parallelization}
   One advantage of the FLOSIC-KLI implementation is that the equations involved are relatively 
   easy to parallelize.
   The most expensive step in the self-consistent FLOSIC calculation is the determination of 
    orbital-dependent potentials, particularly the Coulomb potential, required to compute the SIC terms. 
   However these potentials are independent of each other and can be easily parallelized. 
   The FLOSIC code, which is parallelized using MPI, adds a second level of parallelization for these calculations. 
   The construction of the Hamiltonian using the Jacobi-like method of Yang, Pederson and Perdew\cite{PhysRevA.95.052505}
   is harder to parallelize and creates a bottleneck for larger calculations.
   The present FLOSIC-KLI scheme offers easy parallelization
   and helps in improving scalability.
    In the FLOSIC-KLI approach, the SIC potentials and orbital densities are stored to disk after they are computed.
    Subsequently, each processor reads from file $V^{SIC}$ and $\rho_i$ and 
    the integrals used to generate 
    $M$, $\overline{v}_{i\sigma}^S$, and $\overline{v}_{i\sigma}^{SIC}$ 
    (Eqs. \ref{eq:kli-M}-\ref{eq:kli-vbarSIC})
    are then parallelized over batches of grid points. 
    The contributions from each batch of grid points to the integrals 
    are then reduced across processors.
Construction of the M matrix scales as $O(N_e^2)$ and is thus efficiently parallelized. 
In contrast, the Jacobi-like method scales as $O(N_eN_{b}^3$), where $N_b$ is the number of basis functions in a calculation. 
Since $\rho_i$ which is obtained from the FLO will be localized, we may be able to 
reduce scaling to $O(N_{e})$ by taking advantage of the sparsity of the density.  
In the subsequent section we compare 
the FLOSIC-KLI approach against the FLOSIC-Jacobi approach of Yang, Pederson and 
Perdew\cite{PhysRevA.95.052505} using standard datasets previously employed for assessing the performance
of FLOSIC method. We also report new results on the vertical ionization energies of intermediate 
size water clusters.

\section{Results} \label{sec:results}
\subsection{Atoms: Total energies and Eigenvalues} \label{sec:totalenergies}
\begin{figure}[h]
    \centering
    \includegraphics[width=1.0\columnwidth]{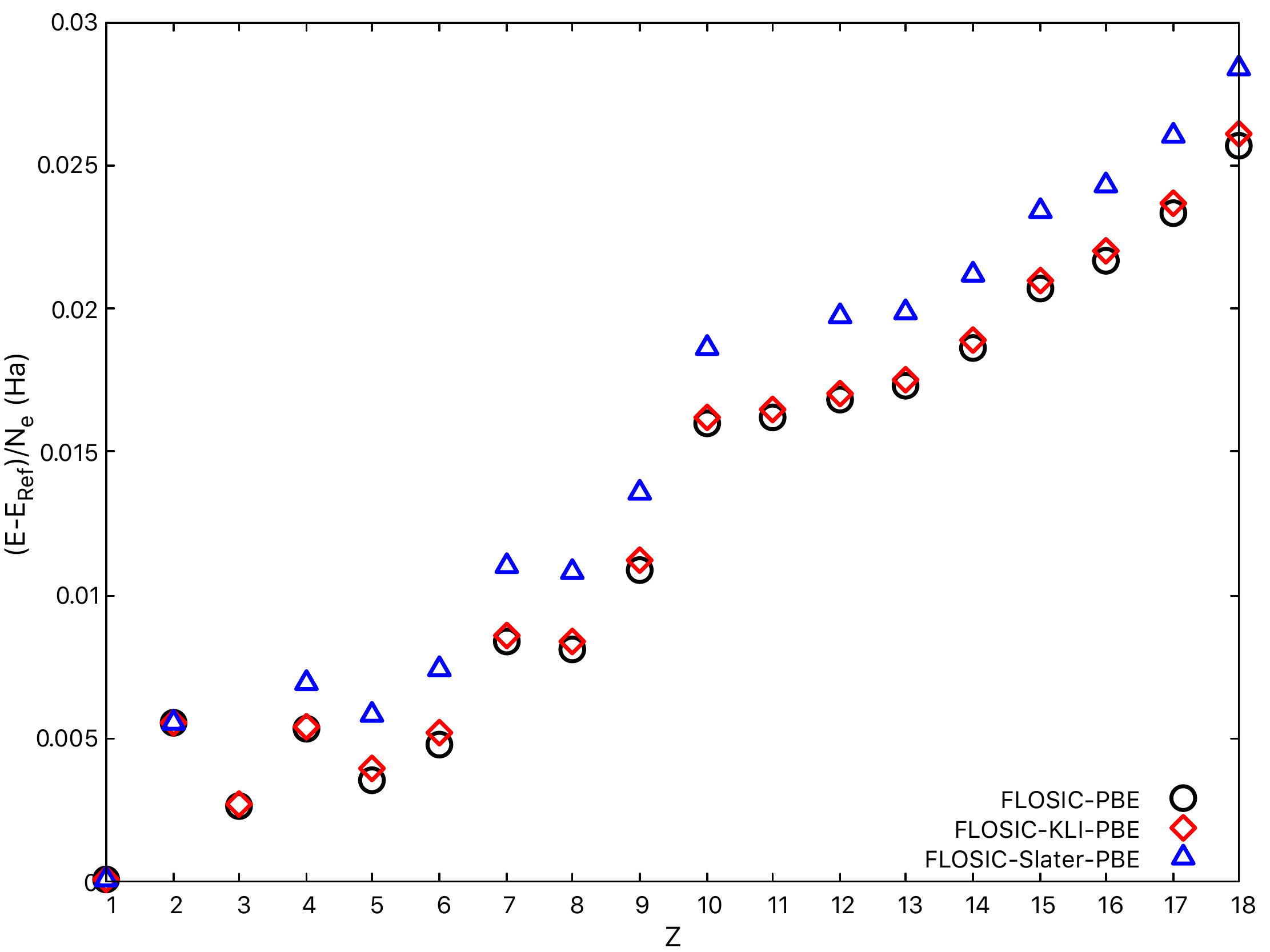}
    \caption{Atomic total energies (in Ha) for FLOSIC-Jacobi (color), FLOSIC-KLI (color), and FLOSIC-Slater (color) compared against the reference values of Ref. [\onlinecite{PhysRevA.47.3649}]. $(E-E_{Ref})/N_e$ is shown, where $N_e$ is the number of electrons.}
    \label{fig:atomicenergies}
\end{figure}

FLOSIC energies for atoms from H-Ar (Z=1-18) are compared against accurate total energies reported by Chakravorty 
et al. \cite{PhysRevA.47.3649}
 We report the deviation on a per electron basis as $(E-E_{Ref})/N_e$, where $E$ is the FLOSIC energy and $E_{Ref}$ is the reference energy.
We find that the FLOSIC-KLI results give very close energies compared with the original FLOSIC implementation, with a mean absolute error (MAE) of 0.161 Ha for FLOSIC-KLI compared to 0.158 Ha for FLOSIC-Jacobi.
The FLOSIC-Slater calculations perform slightly worse in each case, as shown in Fig. \ref{fig:atomicenergies}, and did not converge for the lithium and sodium atoms. FLOSIC-KLI calculations for these atoms were converged by using the $C_\sigma=min(x_{i\sigma})$ factor, as detailed in Sec. \ref{sec:flosickliequations}.
Neglecting these atoms, FLOSIC-KLI, FLOSIC-Jacobi, and FLOSIC-Slater give a MAE of 0.170, 0.167, and 0.192 Ha, respectively.

\begin{figure}[h]
    \centering
    \includegraphics[width=1.0\columnwidth]{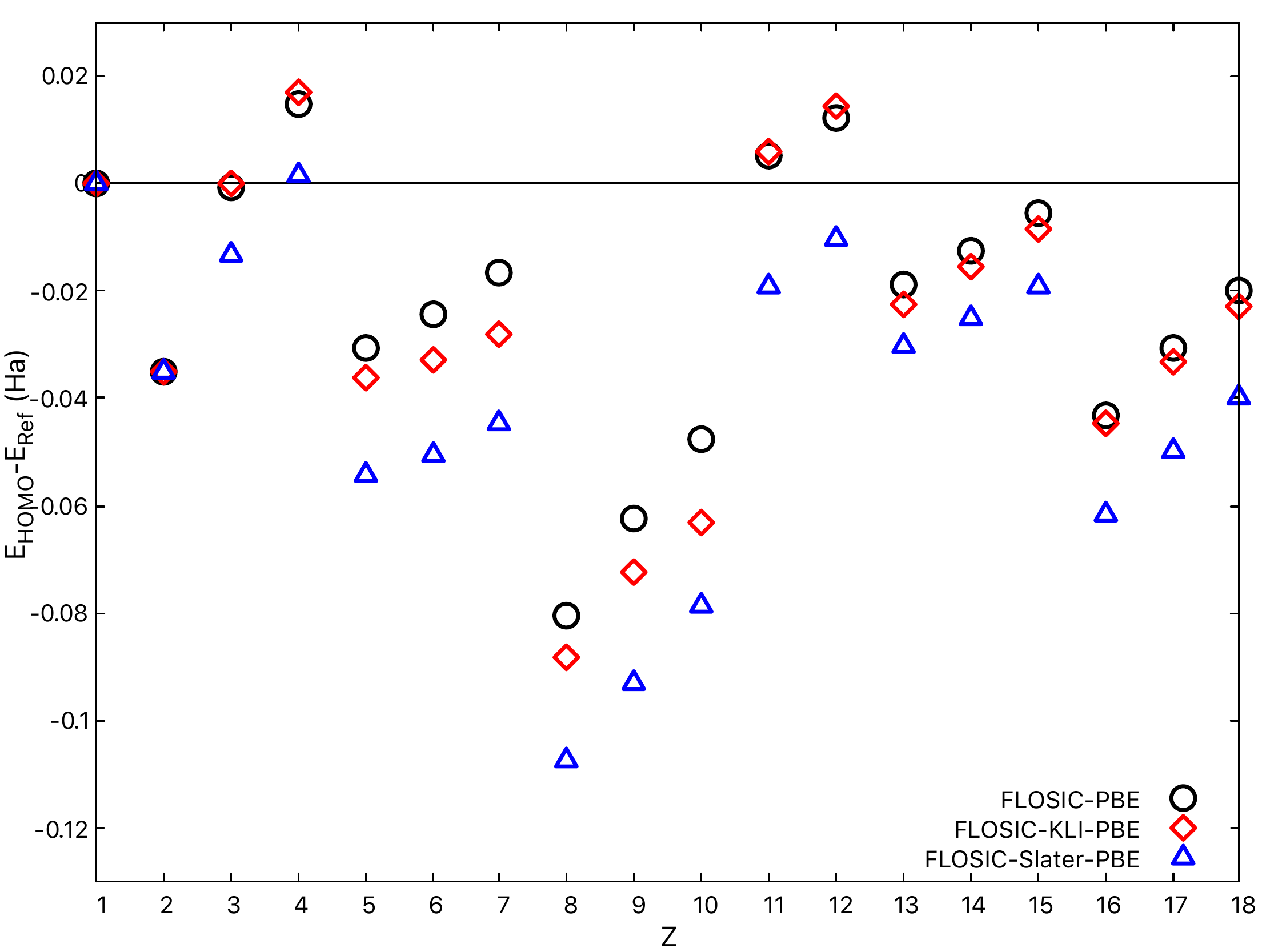}
    \caption{Error in HOMO eigenvalues compared to experimental IPs \cite{NIST_ASD}.}
    \label{fig:evals}
\end{figure}

The vertical ionization potential (vIP) can be obtained from 
the negative of the highest occupied orbital (HOO) eigenvalue.
For the exact exchange-correlation functionals, they are equal\cite{perdew1982density,almbladh1985exact,perdew1997comment}.
For the approximate functionals, the quality of the asymptotic 
behavior of the exchange functionals determines the accuracy 
of the HOO as an approximation to the vIP. All semi-local 
functionals perform poorly in this regard.
In Fig. \ref{fig:evals} we compare the HOO eigenvalues to experimental ionization
potentials (IPs) \cite{NIST_ASD}. Table \ref{table:homo-ip} shows the MAEs and mean absolute relative 
errors (MAREs) for the FLOSIC-Jacobi and FLOSIC-KLI approaches, as well as the less 
accurate FLOSIC-Slater approximation.  These results show good agreement between 
FLOSIC-Jacobi and FLOSIC-KLI, with a difference in MARE of less than 1\%. FLOSIC-Slater performs slightly worse with a MARE 3.8\% higher than that of FLOSIC-Jacobi.

\begin{table}
 \caption{ MAE (in Ha), and MARE (\%) of HOMO eigenvalues compared to experimental IPs \cite{NIST_ASD}. FLOSIC-Jacobi results from [\onlinecite{FLOSICSCANpaper}].  }
 \label{table:homo-ip}
 \begin{tabular}{cccc}
  \hline
 \hline
  & FLOSIC-Jacobi &	FLOSIC-KLI &	FLOSIC-Slater \\
 \hline
 MAE (Ha)   &  0.026 &  0.030  &   0.041   \\
 MARE (\%)  &  5.67 &  6.62   &   9.44    \\
 \hline
 \end{tabular}
\end{table}

\subsection{Atomization energies} \label{sec:atomization}
FLOSIC-Jacobi and FLOSIC-KLI are also used to calculate the total and atomization
energies (AEs) of a set of 37 molecules taken from the G2/97 test set\cite{doi:10.1063/1.460205}. 
In addition, we include the six molecules from the AE6 test set\cite{doi:10.1021/jp035287b}, 
as well as HBr, LiBr, NaBr, FBr, and Br$_2$. Most of the geometries 
were optimized using B3LYP with the 6-31G(2df,p) basis \cite{NIST_CCCBD}. 
The geometries for O$_2$, CO, CO$_2$, C$_2$H$_2$, Li$_2$, CH$_4$, NH$_3$, 
and H$_2$O were optimized using the PBE functional and the default NRLMOL basis set. 
The atomization energy (AE) of a molecule is defined as 
   $AE =  \sum_i^{N_{atoms}} E_i - E_{mol} > 0, $
where $E_i$ is the energy of individual atoms, $N_{atom}$ is 
the number of atoms in the molecule, and $E_{mol}$ is the total energy of the molecule. 
For the $AE6$ set, 
we find that FLOSIC-KLI has slightly larger MARE (7.51\%) compared to 
FLOSIC-Jacobi (6.82\%).

\begin{figure}[h]
    \centering
    \includegraphics[width=1.0\columnwidth]{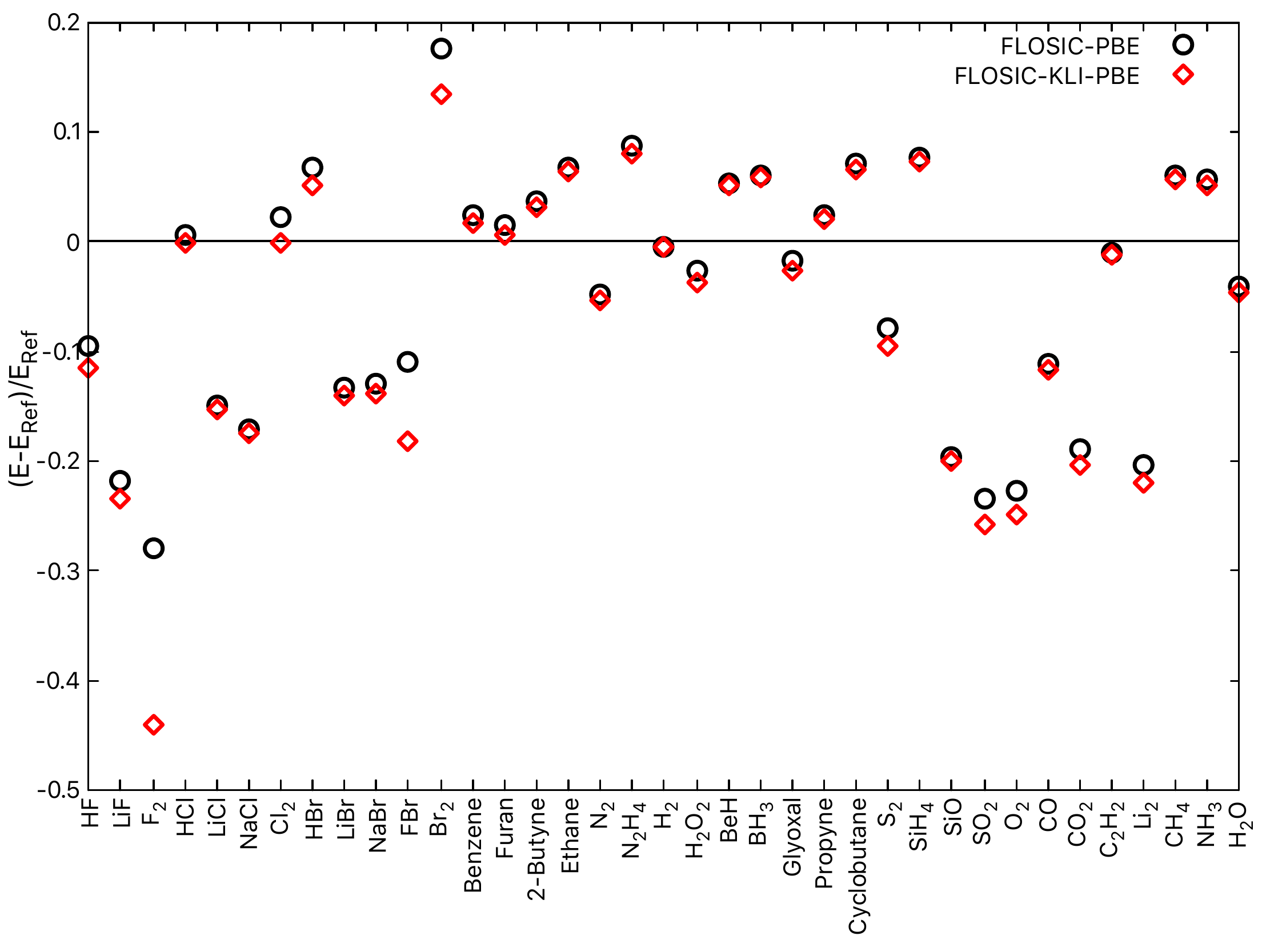}
    \caption{Relative error 
    $(E-E_{Ref})/E_{Ref}$ 
    of atomization energies of molecules compared against the reference experimental values found in Ref. [\onlinecite{NIST_ASD}]. }
    \label{fig:atomizationenergies}
\end{figure}

For the larger set of molecules  the average errors in calculated AEs 
for FLOSIC-Jacobi and FLOSIC-KLI calculations  are summarized in  Table \ref{table:AE}.
Experimental values are taken from Ref. \onlinecite{NIST_ASD}.
The MAREs are 9.67\% and 10.00\% for FLOSIC-Jacobi and FLOSIC-KLI, respectively. 
Figure \ref{fig:atomizationenergies} shows a close agreement between two implementations for most systems, 
except for F$_2$. 
\textcolor{black}{
}

\begin{table}[h]
 \caption{ Atomization energies for the set of molecules featured in Fig. \ref{fig:atomizationenergies}. MAE (kcal/mol) and MARE (\%) are shown.  }
 \label{table:AE}
 \begin{tabular}{ccc}
 \hline
 \hline
  & FLOSIC-Jacobi &	FLOSIC-KLI  \\
 \hline
 MAE (kcal/mol) &  84.29 &  83.32   \\
 MARE (\%)      &   9.67 &   10.00    \\
 \hline
 \end{tabular}
\end{table}

\begin{figure}[h]
    \centering
    \includegraphics[width=1.0\columnwidth]{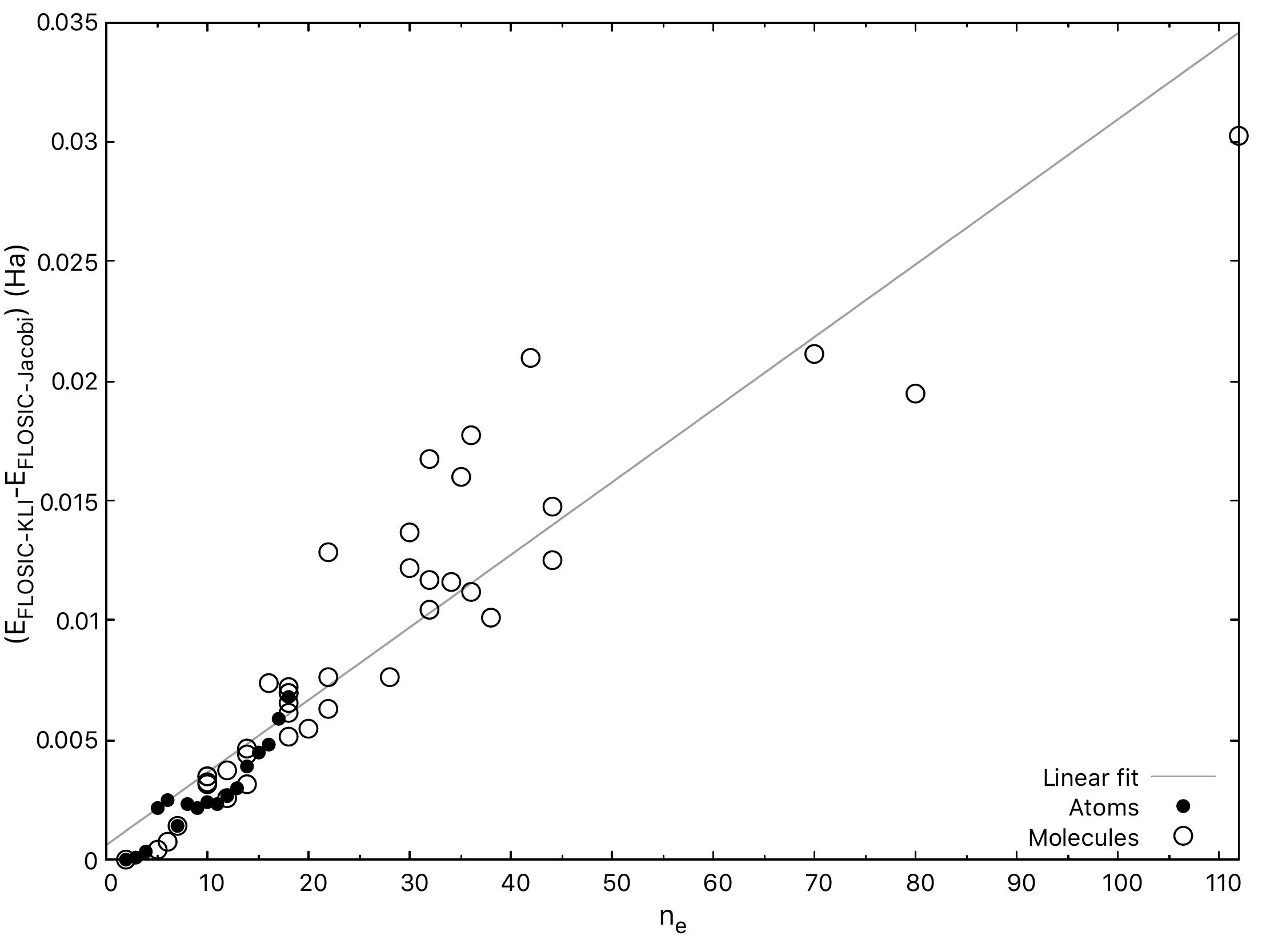}
    \caption{Difference in total energy (in a.u.) between FLOSIC-Jacobi and FLOSIC-KLI calculations as a function of the number of electrons in the system. Linear fit of data shown as solid line.}
    \label{fig:diffenergy}
\end{figure}

Figure \ref{fig:diffenergy} plots the differences in total energies between the FLOSIC-Jacobi and FLOSIC-KLI implementations as a function of number of electrons for all atoms and molecules tested. 
The plot shows a linear behavior, signifying the error per electron to fall within some constant range. When calculating quantities such as AEs, these differences cancel out.

\section{Polarizability of H$_2$ chains } \label{sec:h2-chains}

Most semi-local functionals perform poorly in predicting the response of charge distributions 
to electric fields for molecular chains and polymers\cite{PhysRevLett.83.694,gritsenko2000origin,mori2003accurate,kirtman2008calculation,vargas2014electronic}.
The polarizabilities predicted by semi-local functionals are 
severely overestimated. 
\textcolor{black}{However, recent work by  Aschebrock and K\"ummel shows that meta-GGA functionals
constructed by considering KS potential related properties such as
the derivative discontinuity and its density response can provide an accurate description 
of polarizabilities\cite{PhysRevResearch.1.033082}.
}
The chains of hydrogen molecules have been extensively used as model systems to 
examine performance of DFAs in predicting  the electric response 
of molecular chains\cite{PhysRevLett.83.694,kuemmel2004electrical,sekino2005influence,baer2005density,PhysRevB.75.045101,PhysRevA.77.060502,PhysRevA.77.060502,Maitra2007,korzdorfer2008electrical,messud2009polarizabilities,doi:10.1002/jcc.25586}.
The overestimation of polarizabilities has been understood as a result of a missing field-counteracting
term in the response part of the XC potentials of semi-local functionals\cite{PhysRevLett.83.694,kuemmel2004electrical}.
Here, we use hydrogen chains to examine how well FLOSIC-KLI compares with FLOSIC-Jacobi for the polarizabilities of these systems.
For this purpose we use {\em {finite-field}} method with an electric field of  $h=1.0*10^{-3}$ a.u.
The polarizability is calculated using a second-order central finite difference approach. 
The z-component of the polarizability $\alpha_{zz}$ is calculated as
\begin{equation}
    \alpha_{zz}=\frac{d\mu_z}{dF_z}=\frac{d^2E}{d^2F_z}=\frac{E(-h)-2E(0)+E(h)}{h^2}
\end{equation}
where $h$ is the $z$-component of the electric field.

Table \ref{table:dipole} shows the calculated polarizabilities for H$_n$ chains comparing PBE, FLOSIC-Jacobi and FLOSIC-KLI.
We constructed  linear chains of hydrogen atoms by placing hydrogen atoms 
with  alternating distances of 2 and 3 Bohr.
Initial FODs were generated by placing a spin-up and spin-down FOD at the midpoint between each bonded H$_2$ molecule. 
Polarizabilities were then calculated using the initial guess as well as by optimizing FODs using a $10^{-4}$ Ha/Bohr convergence criterion.
In the case of the H$_{100}$ chain, the FOD positions were not optimized.
Table \ref{table:dipole} shows the polarizabilities calculated using the initial guess of FODs 
show a mean average error of $2.7\%$ compared to the final optimized calculations,
and lie between the FOD-optimized calculations and the MP4 reference calculations.

\begin{table}[h!]
 \caption{ Polarizabilities $\alpha_{zz}$ of H$_2$ chains. MP4 and CCSD values from Ref. [\onlinecite{PhysRevA.77.060502}].
 Mean absolute relative error (MARE) relative to CCSD(T) calculations for H$_{4-12}$.}
 \label{table:dipole}
  \begin{tabular}{ccccccc|c} 
 \hline
 \hline
 
Method~~  & ~ H$_4$  &~  H$_6$  &~  H$_8$  &~  H$_{12}$  & ~ H$_{14}$  ~&  H$_{100}$ ~& MARE(\%)\\
\hline
PBE  &  36.0  &  69.1  &  108.4  &  197.0  &  243.9  &  2,600.1 & 43.1\\
FLOSIC-KLI\footnote{FOD positions in these calculations are not optimized.}
               &  32.1  &  56.8  &  83.6  &  158.7  &  173.7  &  1,417.7 & 17.3 \\
FLOSIC-KLI     &  32.1  &  59.2  &  88.6  &  158.5  &  180.5  &   &  20.1\\
FLOSIC-Jacobi  &  31.2  &  60.3  &  90.5  &  156.9  &  194.8  &   & 20.3 \\
MP4  
               &  29.5  &  51.9  &  75.2  &  127.3  &  155.0  & & 3.3 \\
CCSD(T) 
               &  28.7  &  50.2  &  73.4  &  122.0  &         & &   \\
 \hline
 \end{tabular}
\end{table}

\begin{figure}[h]
    \centering
    \includegraphics[width=1.0\columnwidth]{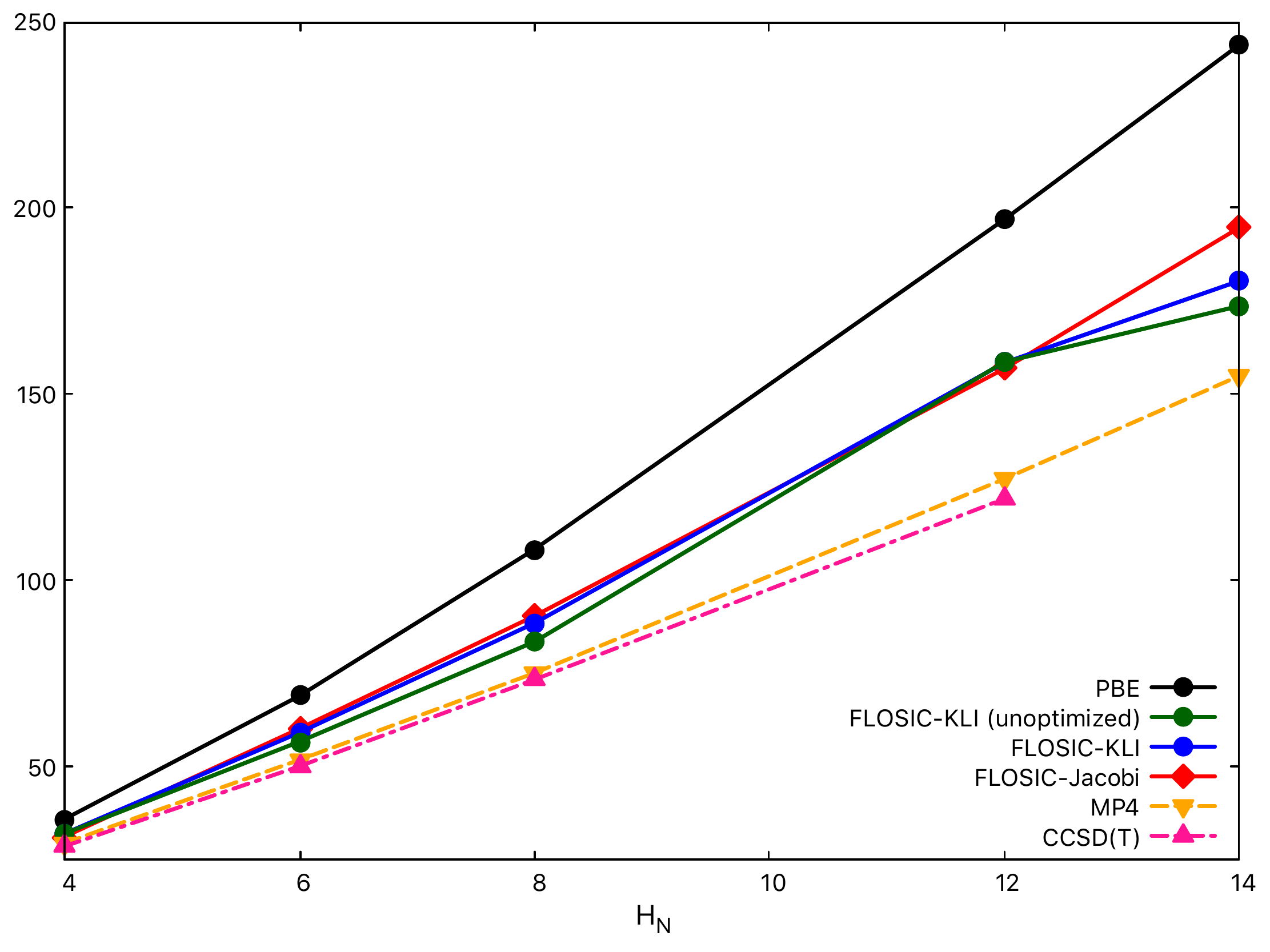}
    \caption{Polarizabilities $\alpha_{zz}$ of H$_2$ chains plotted as a function of number of hydrogen atoms. MP4 and CCSD values from Ref. [\onlinecite{PhysRevA.77.060502}].}
    \label{fig:polar}
\end{figure}

\section{HOMO-LUMO Gaps}
\begin{table*}[t!]
 \caption{ HOMO and LUMO eigenvalues for Jacobi and KLI in eV along with the difference in HOMO-LUMO Gaps. The negative 
 of the HOMO eigenvalues approximate the vertical  ionization potentials.}
 \label{table:evals}
  \begin{tabular}{ccc @{\extracolsep{8pt}} ccc}
 \hline
 \hline
  {}&\multicolumn{2}{c}{HOMO (eV)} & \multicolumn{2}{c}{LUMO (eV)} & \\
   \cline{2-3} \cline{4-5}

Molecule  &  FLOSIC-Jacobi  &  FLOSIC-KLI  &  FLOSIC-Jacobi  &  FLOSIC-KLI  &  Gap Difference \\
 \hline
 HF  &  -17.82  &  -17.55  &  -0.53  &  -6.28  &  -6.02  \\
LiF  &  -13.20  &  -13.44  &  -1.32  &  -5.28  &  -3.72  \\
HCl  &  -13.43  &  -13.47  &  -0.90  &  -5.59  &  -4.66  \\
LiCl  &  -10.69  &  -10.72  &  -1.75  &  -4.74  &  -2.97  \\
NaCl  &  -10.06  &  -10.05  &  -2.12  &  -5.33  &  -3.21  \\
Cl$_2$  &  -12.89  &  -12.31  &  -4.68  &  -9.16  &  -5.06  \\
HBr  &  -12.19  &  -12.11  &  -1.38  &  -5.57  &  -4.27  \\
LiBr  &  -9.84  &  -9.85  &  -1.86  &  -4.64  &  -2.77  \\
BrF  &  -12.70  &  -12.44  &  -4.76  &  -9.66  &  -5.15  \\
Br$_2$  &  -11.67  &  -11.19  &  -4.92  &  -8.84  &  -4.40  \\
Benzene  &  -9.08  &  -8.82  &  -1.40  &  -3.63  &  -2.49  \\
Furan  &  -10.36  &  -10.64  &  -0.92  &  -5.65  &  -4.44  \\
2-Butyne  &  -11.00  &  -10.98  &  0.04  &  -4.12  &  -4.18  \\
C$_2$H$_6$  &  -14.30  &  -14.19  &  0.10  &  -4.67  &  -4.88  \\
C$_5$H$_5$  &  -1.62  &  -1.29  &  4.43  &  3.29  &  -1.48  \\
CN$^-$  &  -5.17  &  -4.79  &  6.78  &  2.11  &  -5.06  \\
N$_2$  &  -17.24  &  -16.15  &  -2.05  &  -7.44  &  -6.49  \\
BH$_3$  &  -14.36  &  -14.48  &  -3.01  &  -8.75  &  -5.63  \\
Cyclobutane  &  -13.10  &  -13.08  &  0.12  &  -4.51  &  -4.65  \\
S$_2$  &  -10.94  &  -11.08  &  -4.65  &  -7.80  &  -3.01  \\
SiH$_4$  &  -13.99  &  -14.01  &  0.17  &  -4.91  &  -5.05  \\
SiO  &  -12.41  &  -12.04  &  -2.92  &  -6.89  &  -4.34  \\
SO$_2$  &  -14.48  &  -14.19  &  -4.64  &  -9.99  &  -5.64  \\
 \hline
 \end{tabular}
\end{table*}

\begin{figure}[h]
    \centering
    \includegraphics[width=1.0\columnwidth]{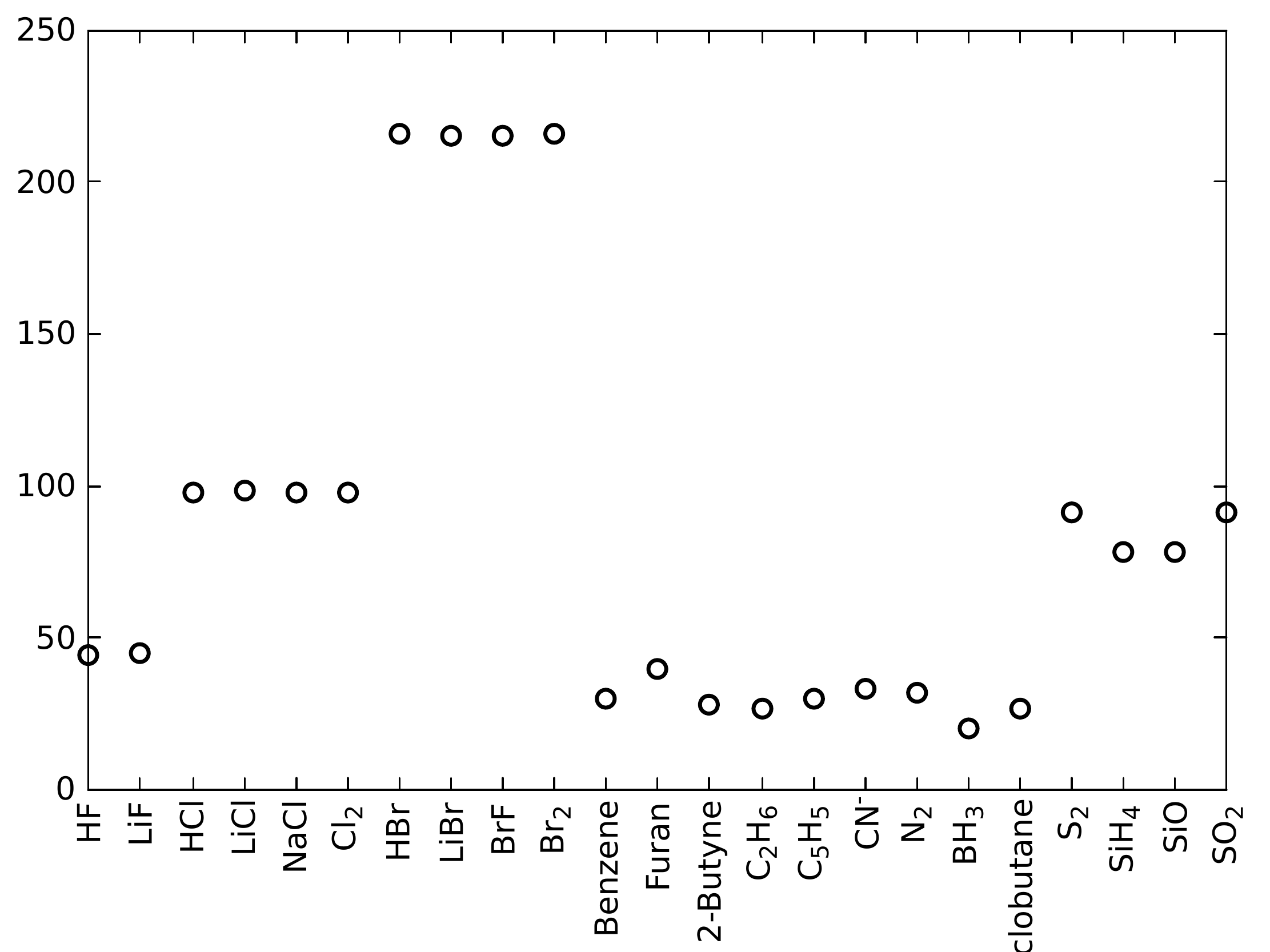}
    \caption{The difference between the width of occupied 
    eigenvalue spectrum (i.e., the difference in deepest 1s Core eigenvalue and HOMO eigenvalue) (in eV) between FLOSIC-Jacobi and FLOSIC-KLI calculations.}
    \label{fig:corehomo}
\end{figure}

There has been considerable discussion 
about the interpretation of Kohn-Sham orbital energies as electron 
removal energies or the differences between the orbital energies 
as the excitation energies\cite{ Janak1978, perdew1982density,almbladh1985exact,PhysRevA.30.2745,perdew1997comment, seidl1996generalized, Politzer1998, Stowasser1999, Chong2002, Gruning2006, PhysRevB.77.115123, Teale2008, PhysRevLett.100.146401, Yang2012, kronik2012excitation, Baerends2013, VanMeer2014, Perdew2017, Perdew2018, doi:10.1063/1.5026951, Baerends2020, Godby1989,
PhysRevLett.100.146401, Yang2012, kronik2012excitation, Baerends2013, VanMeer2014, Perdew2017, Perdew2018, doi:10.1063/1.5026951, Baerends2020, Godby1989}.
Despite these, the density of states from Kohn-Sham calculations
are often used to interpret experimental observations.
DFAs that have explicit orbital dependence,
such as hybrid or meta-GGA functionals, 
are typically implemented using the generalized Kohn-Sham scheme\cite{seidl1996generalized}.
 The self-consistent implementation of the PZ-SIC method
 using the Jacobi scheme (FLOSIC-Jacobi) 
 is like the generalized KS scheme used 
 for hybrid DFAs or meta-GGAs. 
The FLOSIC-KLI method gives a multiplicative effective potential as in the 
standard KS scheme. As seen in previous sections, these two implementations 
of the PZ-SIC give total 
atomic energies, atomization energies and polarizabilities that agree within 1-2\% .
The eigenvalues, especially the eigenvalues of the 
unoccupied molecular orbitals (LUMOs), in the two approaches 
are however very different. 
The FLOSIC-Jacobi LUMOs are essentially same as that of the uncorrected 
functional as the Jacobi scheme does not affect the unoccupied orbitals.
As can be seen from Table \ref{table:evals} and Fig. \ref{fig:corehomo}, the FLOSIC-KLI LUMO  (and 
higher unoccupied orbitals) are substantially lowered compared 
to the FLOSIC-Jacobi LUMO. The calculations also show that the eigenvalues of the 
core orbitals (especially those of 1s orbitals) are destabilized by several eV in 
the FLOSIC-KLI scheme. Since the HOMO eigenvalues between the FLOSIC-Jacobi 
and FLOSIC-KLI agree within 1\%,
the eigen-spectrum in the FLOSIC-KLI scheme is compressed compared to
FLOSIC-Jacobi. This behavior is illustrated in Fig. \ref{fig:corehomo} which shows the
difference between the first (lowest) and the highest occupied eigenvalues
in the FLOSIC-Jacobi and FLOSIC-KLI methods. This means the core electron 
binding energies if estimated from the absolute eigenvalues of core 
electrons will differ by several eVs in the two approaches. This 
would also lead to differences in the prediction 
of the core-valence excitations used in simulating 
near-edge x-ray absorption fine structure (NEXAFS) spectra
by two approaches.

\begin{table}[h]
    \caption{HOMO and LUMO eigenvalues and HOMO-LUMO gaps in eV for water clusters calculated using FLOSIC-KLI.}
    \centering
    \begin{tabular}{ccc @{\extracolsep{8pt}}cc @{\extracolsep{8pt}} cc}
     \hline
    \hline
    {}&\multicolumn{2}{c}{HOMO (eV)} & \multicolumn{2}{c}{LUMO (eV)} & \multicolumn{2}{c}{Gap (eV)} \\
    \cline{2-3} \cline{4-5} \cline{6-7}
H$_2$O Molecules  &  LDA  &  PBE  &  LDA  &  PBE  &  LDA  &  PBE \\ 
\hline
1  &  -14.75  &  -14.27  &  -6.30  &  -5.88  &  8.45  &  8.39 \\ 
5  &  -14.48  &  -13.95  &  -6.40  &  -5.87  &  8.08  &  8.07 \\ 
10  &  -14.11  &  -13.60  &  -6.84  &  -6.21  &  7.27  &  7.39 \\ 
15  &  -14.07  &  -13.56  &  -7.27  &  -6.59  &  6.79  &  6.96 \\ 
20  &  -14.49  &  -13.91  &  -7.03  &  -6.30  &  7.46  &  7.61 \\ 
21  &  -13.82  &  -13.31  &  -7.04  &  -6.35  &  6.78  &  6.96 \\ 
22  &  -14.44  &  -13.91  &  -7.17  &  -6.47  &  7.27  &  7.44 \\ 
23  &  -13.97  &  -13.49  &  -7.12  &  -6.44  &  6.85  &  7.05 \\ 
24  &  -14.24  &  -13.74  &  -7.21  &  -6.51  &  7.03  &  7.23 \\ 
25  &  -14.17  &  -13.63  &  -7.01  &  -6.28  &  7.16  &  7.35 \\ 
26  &  -14.08  &  -13.56  &  -7.18  &  -6.46  &  6.90  &  7.10 \\ 
27  &  -14.23  &  -13.75  &  -7.21  &  -6.51  &  7.02  &  7.24 \\ 
28  &  -14.25  &  -13.71  &  -7.27  &  -6.51  &  6.98  &  7.20 \\ 
29  &  -14.21  &  -13.68  &  -7.24  &  -6.50  &  6.97  &  7.18 \\ 
30  &  -13.97  &  -13.64  &  -7.45  &  -6.70  &  6.51  &  6.95 \\ 
    \hline
    \end{tabular}
    \label{tab:waterip}
\end{table}

\section{Ionization potentials of water clusters}
 We have used the present methodologies to obtain the vertical
 ionization potentials of water clusters from  (H$_2$O)$_{21}$ to  (H$_2$O)$_{30}$. 
 The geometries of these clusters are from the recent study by Rakshit\cite{rakshit2019atlas}
 and coworkers. These authors performed a large scale search for the 
 putative minima of water clusters using Monte Carlo basin
 paving approach with a polarizable Thole-Type model for force field.
 These geometries were further refined at the MP2/aug-cc-pVTZ level of theory.
 The FLOSIC-KLI calculations were performed on the most stable
 water clusters at MP2/aug-cc-pVTZ level. The FODs for these
 clusters were obtained using the fodMC code\cite{Schwalbe2019}.
 No further optimizations of FODs were performed. To examine 
 how well this approach  works for the properties of water clusters studied herein,
 we optimized the FODs using FLOSIC code for the (H$_2$O)$_{20}$ cluster. We find that 
 the forces on the FODs are very small and the optimization changes 
 the HOMO eigenvalue by 0.4\%.
 The HOMO and the LUMO eigenvalues of water clusters along with HOMO-LUMO gap 
 are presented in Table \ref{tab:waterip}. The vertical ionization potentials
 are the absolute values of the HOMO eigenvalues. The ionization potentials 
 of (H$_2$O)$_{21}$-(H$_2$O)$_{30}$ water clusters are in the range 13.8 eV 
 to 14.4 eV and do not show systematic variation with respect to size. 
 Recently, Akter and  coworkers\cite{Akter2020} studied small water clusters using 
 PZSIC and locally scaled self-interaction methods. They found that the 
 vertical ionization potentials obtained as an absolute 
 of the HOMO eigenvalue within the FLOSIC-LSDA scheme show systematic 
 overestimation of approximately 2 eV when compared with CCSD(T) ionization 
 potentials. By adding this shift, FLOSIC-KLI ionization potentials 
 are in good agreement with CCSD(T) energies.
 Likewise, the PBE FLOSIC-KLI HOMO-LUMO 
 gaps are in the range of 6.7 eV to 7.6 eV. For the water molecule
 the HOMO-LUMO gap is  8.39 eV. Thus there is  about 1 to 1.4 eV reduction
 of the HOMO-LUMO gap from water molecule to water clusters containing 20-30
 water molecules.

 \comment{
 \textcolor{black}{To do:
 Core electron binding energies compared to FLOSIC-Jacobi and FLOSIC-KLI eigenvalues in Table \ref{tab:cebe}. Overestimated in FLOSIC-Jacobi, underestimated in FLOSIC-KLI.}
\begin{table}[h]
    \centering
    \begin{tabular}{cccc}
    \hline
Molecule & Expt. & FLOSIC-Jacobi & FLOSIC-KLI \\
    \hline
HF    & 694.23 & -21.17 & 23.26 \\
F$_2$ & 696.69 & -21.99 & 24.64 \\
N$_2$ & 409.98 & -16.40 & 16.74 \\
    \hline
    \end{tabular}
    \caption{Core 1s core excitation binding energies (in eV) and error from FLOSIC-Jacobi and FLOSIC-KLI eigenvalues}
    \label{tab:cebe}
\end{table}
 }

\section{Conclusion}
To summarize, we have implemented the FLOSIC method  using the optimized 
effective potentials  with the Krieger-Li-Iafrate (KLI) approximation.
The implementation was tested by computing 
the atomic  energies, atomization energies, the eigenvalues and the ionization 
potentials using standard data sets, polarizabilities of hydrogen chains  
and comparing the results with those obtained using 
the FLOSIC-Jacobi method of Yang, Pederson and Perdew\cite{PhysRevA.95.052505}. 
It is found that the FLOSIC-KLI  approach gives results 
that are in close agreement within 1-2\% 
of the FLOSIC-Jacobi method. 
We have also used the FLOSIC-KLI scheme to predict the vertical ionization energies of 
water clusters.

The FLOSIC-KLI is a desirable approach for larger calculations as it allows more 
efficient and scalable parallelization than the FLOSIC-Jacobi method. 
Another desirable feature of FLOSIC-KLI approach is that 
it provides self-interaction 
corrected virtual orbitals. The virtual orbitals are required for the calculation 
of excitation energies using the  time-dependent density functional or for magnetic 
anisotropy calculations using the Pederson-Khanna method\cite{PhysRevB.60.9566}. 
Such applications will be investigated in the future.

\section{Acknowledgements}
Authors acknowledge Profs. Mark Pederson, Koblar Jackson and 
Yoh Yamamoto for reading the manuscript and helpful comments.
Authors  acknowledge support by the US Department of Energy, Office of
Science, Office of Basic Energy Sciences, as part of the Computational Chemical Sciences
Program under Award No. DE-SC0018331.

\section{Data availability}
The data that support the findings of this study are available from the corresponding author upon reasonable request.

\bibliography{bibtex_common_kli,bibnew}

\end{document}